%% file: paper.tex
\documentclass[a4paper,10pt,journal]{IEEEtran}
\IEEEoverridecommandlockouts 

\makeatletter
\newcommand{\AddInputPath}[1]{%
  \ifx\input@path\@undefined
    \def\input@path{#1}
  \else
    \g@addto@macro{\input@path}{#1}
  \fi
}
\makeatother
\AddInputPath{{../}}

\usepackage{etex}

\usepackage[caption=false,font=footnotesize]{subfig} 

\usepackage{relsize}

\usepackage{array}
\usepackage{booktabs,tabularx}
\usepackage{multirow}
\usepackage[binary-units=true,per-mode=symbol,detect-mode=true]{siunitx}
\usepackage{graphicx}
\graphicspath{ {images/} }
\usepackage{float}

\usepackage[T1]{fontenc}
\usepackage{textcomp}
\usepackage[utf8]{inputenc}
\usepackage[final]{microtype}
\usepackage{icomma}
\usepackage{xspace}

\usepackage[tbtags]{amsmath}
\usepackage{amssymb,amsfonts,bm}
\usepackage{mathtools} 
\usepackage{dsfont}
\usepackage{mathrsfs}
\usepackage{accents}
\usepackage{empheq}
\usepackage{nccmath}
\usepackage{balance}

\usepackage{color}
\usepackage{calc}
\usepackage{tikz}
\usepackage{pgfplots,pgfplotstable}

\usepackage{pdftexcmds}
\makeatletter
\newcommand{\strequal}[2]{\pdf@strcmp{#1}{#2}==0}
\makeatother

\usepackage[capitalize]{cleveref}
\usepackage{refcount}

\usepackage[inline]{enumitem}
\usepackage{algorithm}
\usepackage{algpseudocode}
\makeatletter
\newcommand{\algmargin}{\the\ALG@thistlm}
\makeatother
\newlength{\whilewidth}
\algdef{SE}[parWHILE]{parWhile}{EndparWhile}[1]
  {\parbox[t]{\dimexpr\linewidth-\algmargin}{%
     \hangindent\whilewidth\strut\algorithmicwhile\ #1\ \algorithmicdo\strut}}{\algorithmicend\ \algorithmicwhile}%
\algnewcommand{\parState}[1]{\State%
  \parbox[t]{\dimexpr\linewidth-\algmargin}{\strut #1\strut}}

\makeatletter
\newcommand\fs@spaceruled{\def\@fs@cfont{\bfseries}\let\@fs@capt\floatc@ruled
  \def\@fs@pre{\vspace{.05in}\hrule height.8pt depth0pt \kern2pt}%
  \def\@fs@post{\kern2pt\hrule\relax}%
  \def\@fs@mid{\kern2pt\hrule\kern2pt}%
  \let\@fs@iftopcapt\iftrue}
\makeatother

\usepackage{glossaries}
\usepackage{ifthen}
\usepackage{cite}
\usepackage{multibib}

\usepackage{comment}
\usepackage{todonotes}
\let\legacytodo\todo
\newcommand{\ruggedtodo}[2][]{\tikzexternaldisable\legacytodo[#1]{#2}\tikzexternalenable}
\renewcommand{\todo}[1]{\ruggedtodo[inline]{#1}}

\input{acronyms}
\glsenableentrycount
\makeglossaries

\pgfplotsset{compat=1.17}
\usetikzlibrary{positioning}
\usetikzlibrary{calc}
\usetikzlibrary{math}
\usetikzlibrary{fit}
\usetikzlibrary{intersections}
\usetikzlibrary{decorations.pathreplacing}
\usetikzlibrary{decorations.markings}
\usetikzlibrary{3d,angles}
\usetikzlibrary{spy}

\usetikzlibrary{external}

\tikzset{
	antenna/.pic={
		\draw[thick] (0,0) -- ++(120:2mm) -- ++(0:2mm) -- cycle -- (0,-1.5mm);
	}
}

\makeatletter
\newcommand\transformxdimension[1]{
    \pgfmathparse{((#1/\pgfplots@x@veclength)+\pgfplots@data@scale@trafo@SHIFT@x)/10^\pgfplots@data@scale@trafo@EXPONENT@x}
}
\newcommand\transformydimension[1]{
    \pgfmathparse{((#1/\pgfplots@y@veclength)+\pgfplots@data@scale@trafo@SHIFT@y)/10^\pgfplots@data@scale@trafo@EXPONENT@y}
}
\makeatother

\crefname{equation}{}{}
\crefrangeformat{equation}{(#3#1#4)--(#5#2#6)}

\undef\mod
\DeclareMathOperator\mod{mod}

\newcommand{\norm}[1]{\ensuremath{\left\lVert #1 \right\rVert}}

\let\card=\abs
\let\vec\bm

\allowdisplaybreaks[1]

\DeclareSIUnit \dBm {dBm}
\DeclareSIUnit \dBW {dBW}
\DeclareSIUnit \bpcu {bpcu}

\usepackage{etoolbox}
\AtBeginEnvironment{equation}{\small}
\AtEndEnvironment{equation}{\normalfont}
\AtBeginEnvironment{gather}{\small}
\AtEndEnvironment{gather}{\normalfont}
\AtBeginEnvironment{split}{\small}
\AtEndEnvironment{split}{\normalfont}
\AtBeginEnvironment{equation*}{\small}
\AtEndEnvironment{equation*}{\normalfont}
\AtBeginEnvironment{align}{\small}
\AtEndEnvironment{align}{\normalfont}
\AtBeginEnvironment{align*}{\small}
\AtEndEnvironment{align*}{\normalfont}
\AtBeginEnvironment{alignat}{\small}
\AtEndEnvironment{alignat}{\normalfont}
\AtBeginEnvironment{flalign}{\small}
\AtEndEnvironment{flalign}{\normalfont}
\AtBeginEnvironment{multline}{\small}
\AtEndEnvironment{multline}{\normalfont}
\AtBeginEnvironment{multline*}{\small}
\AtEndEnvironment{multline*}{\normalfont}

\AtBeginEnvironment{algorithmic}{\footnotesize}
\makeatletter
\newcommand\floatc@ruledn[2]{\small{\@fs@cfont #1} #2\par}
\newcommand\fs@rulednew{\fs@ruled\let\@fs@capt\floatc@ruledn}
 \makeatother
\floatstyle{rulednew}
\restylefloat{algorithm}

\DeclareFontFamily{U}{mathx}{\hyphenchar\font45}
\DeclareFontShape{U}{mathx}{m}{n}{
      <5> <6> <7> <8> <9> <10>
      <10.95> <12> <14.4> <17.28> <20.74> <24.88>
      mathx10
      }{}
\DeclareSymbolFont{mathx}{U}{mathx}{m}{n}
\DeclareMathSymbol{\bigtimes}{1}{mathx}{"91}

\pgfplotscreateplotcyclelist{default}{%
	blue,mark=*\\%
	red,mark=star\\%
	teal,mark=square*\\%
	brown!60!black,mark=otimes*\\%
}

\pgfplotsset{
	accuracyplot/.style={
			thick,
			xlabel={Time [$\si{\hour}$]},
			ylabel={Top-1 Accuracy},
			ylabel near ticks,
			grid=major,
			minor x tick num = 3,
			minor y tick num = 1,
			xmin = 0,
			ymin = 0,
			ymax = .9,
			legend pos=south east,
			legend cell align=left,
			legend style={font=\footnotesize},
			legend entries = {FedSat, FedSat+, FedAsync, FedAvg},
			cycle list name=default,
			no markers,
			width=\axisdefaultwidth,
			height=.60*\axisdefaultheight,
		},
	shorttime/.style={
		xtick = {0,4,...,24},
		xmax = 24,
	},
	longtime/.style={
		xmax = 82,
		xtick = {0,24,...,72},
	}
}

\makeatletter
\newif\ifhbonecolumn
\@ifclasswith{IEEEtran}{onecolumn}{\hbonecolumntrue}{\hbonecolumnfalse}
\makeatother

\begin{document}
\bstctlcite{IEEEexample:BSTcontrol}
\title{Ground-Assisted Federated Learning in\\ LEO Satellite Constellations}

\author{Nasrin~Razmi,~\IEEEmembership{Student Member,~IEEE},
	Bho~Matthiesen,~\IEEEmembership{Member,~IEEE},\\
	Armin~Dekorsy,~\IEEEmembership{Senior~Member,~IEEE},
	and~Petar~Popovski,~\IEEEmembership{Fellow,~IEEE}%
		\thanks{
			N.~Razmi, B.~Matthiesen, and A.~Dekorsy are with the Gauss-Olbers Center, c/o University of Bremen, and the Department of Communications Engineering, University of Bremen, 28359 Bremen, Germany (e-mail: \{razmi,matthiesen,dekorsy\}@ant.uni-bremen.de).
			P.~Popovski is with the Department of Electronic Systems, Aalborg University, 9100 Aalborg, Denmark (e-mail: petarp@es.aau.dk). P.~Popovski is also holder of the U~Bremen Excellence Chair in the Department of Communications Engineering, University of Bremen, 28359 Bremen, Germany.
		}%
		\thanks{
			This work is supported in part by the German Research Foundation (DFG)
			under Germany's Excellence Strategy (EXC 2077 at University of Bremen, University Allowance)
			and by the North-German Supercomputing Alliance.
		}
	}

\maketitle

\begin{abstract}
	In Low Earth Orbit (LEO) mega constellations, there are relevant use cases, such as inference based on satellite imaging, in which 
	a large number of satellites collaboratively train a machine learning model without sharing their local datasets. To address this problem, we propose a new set of algorithms based on Federated learning (FL), including a novel asynchronous FL procedure based on FedAvg that exhibits better robustness against heterogeneous scenarios than the state-of-the-art.
	Extensive numerical evaluations based on MNIST and CIFAR-10 datasets highlight the fast convergence speed and excellent asymptotic test accuracy of the proposed method.
\end{abstract}

\glsresetall


\section{Introduction} \label{sec:intro}

Constellations of small satellites flying in \cgls{leo} are a cost-efficient and versatile alternative to traditional big satellites in medium Earth and geostationary orbits. Several of these constellations are currently deployed with the goal of providing ubiquitous connectivity and low latency Internet service \cite{Portillo2019}. Their integration into terrestrial mobile networks is an active research area, covering various use cases such as Earth observation missions\cite{satmagazine,Qian2020,Kodheli2021,Di2019,Lin2021}.
Presumably, \cgls{ml} will become an essential tool to manage these constellations and utilize their sensor measurements \cite{Vazquez2021,Giuffrida2020,MateoGarcia2021}.

The traditional approach to \cgls{ml} is to aggregate all data in a central location and then solve the learning problem. Considering the vast amounts of data necessary to train deep neural networks \cite{bengio2016}, this involves high transmission costs and delays. Moreover, considering the emergence of variety of private owners of small satellites, it might be prohibited to share the data due to privacy or data ownership concerns. The obvious solution to this dilemma is to train locally and aggregate the derived model parameters only. This is achieved by solving the \cgls{ml} problem collaboratively and only sharing updated model parameters. The distributed \cgls{ml} paradigm taking data heterogeneity and limited connectivity into account is known as \cgls{fl} \cite{Konecny2015,McMahan2017}.
Applying distributed \cgls{ml} to satellite constellations is only natural when considering the general trend towards edge computing \cite{Wang2020a}. For example, in ESA's PhiSat-1 mission, raw \cgls{eo} data is pre-processed using deep \cgls{ml} models on the satellites and only relevant information is transmitted to the ground \cite{Giuffrida2020,MateoGarcia2021}. Since the raw data remains on the satellites, improving the employed \cgls{ml} models based on new observations requires on-board re-training.

A core assumption of the general \cgls{fl} setting is intermittent and unpredictable participation of the clients, i.e., the satellites in the considered scenario. In order to cope with that, asynchronous algorithms have been proposed recently \cite{Xie2020}. However, the distinctive feature of the \cgls{leo} learning scenario is the predictable availability of clients combined with very long periods between visits to the same \cgls{gs}. In this paper, we investigate how this predictive availability impacts the \cgls{fl} scenario when the training process is orchestrated by a \cgls{gs} and propose a novel asynchronous algorithm. We conclusively show that our approach leads to superior training performance when compared to state-of-the-art \cgls{fl} algorithms.
In particular, our key contributions are that we
	\begin{enumerate*}[itemjoin={{; }}, itemjoin*={{, and }}, before=\unskip{: }]
		\item define the \cgls{leo} \cgls{fl} scenario and identify core challenges compared to conventional \cgls{fl}
		\item propose an algorithmic framework and communication protocol for satellite \cgls{fl}
		\item adapt FedAvg \cite{McMahan2017} and FedAsync \cite{Xie2020} to this scenario and propose a novel asynchronous variant of FedAvg that is particularly well suited for ground-assisted \cgls{fl} in satellite constellations 
		\item numerically evaluate the discussed algorithms to verify our theoretical considerations. These results show that the proposed asynchronous \cgls{fl} algorithm has higher robustness against heterogeneous scenarios than FedAsync.
	\end{enumerate*}

\section{System Model and Background on FL} \label{sec:sysmod}
Consider a \cgls{leo} constellation of $K$ satellites in $L$ orbital planes.
In an \cgls{eci} coordinate system, satellite $k$, $k\in\mathcal K= \{1, \dots, K\}$, has trajectory $\vec r_k(t)$ and the \cgls{gs}, although fixed in a constant location on Earth, has trajectory $\vec r_g(t)$.
%
A \cgls{gsl} is feasible if satellite $k$ is visible from the \cgls{gs} at a minimum elevation angle $\alpha_e$, i.e.,
 $\frac{\pi}{2} - \angle (\vec r_g(t), \vec r_k(t) - \vec r_g(t)) \ge \alpha_e$.
In general, only a subset of satellites is connected to the \cgls{gs} at once and the time between contacts is much longer than the actual online time.

Satellite $k$ collects data from its on-board instruments and stores it in a dataset $\mathcal D_k$. Due to different orbits and orbital positions, the datasets of two distinct satellites are disjunct and possibly non-IID. After the data collection phase, the satellites
collaboratively solve an optimization problem of the form
\begin{equation}
	\min_{\vec \theta\in\mathds R^d} \frac{1}{n} \sum_{\vec x \in \mathcal D} f(\vec x; \vec \theta)
	= \min_{\vec\theta\in\mathds R^d} \sum_{k\in\mathcal K}  \frac{n_k}{n} \sum_{\vec x \in \mathcal D_k} \frac{1}{n_k} f(\vec x; \vec \theta) \label{opt}
\end{equation}
with the goal of training a machine learning model,
where $\mathcal D = \bigcup_{k\in\mathcal K} \mathcal D_k \subset \mathds R^m$, $n_k = \card{\mathcal D_k}$, and $n = \sum_{k\in\mathcal K} n_k$.
The objective $\frac{1}{n} \sum_{\vec x \in \mathcal D} f(\vec x; \vec \theta)$ is an empirical loss function defined by the training task, where $f(\vec x; \vec\theta)$ is the training loss for a data point $\vec x\in\mathcal D$ and model parameters $\vec \theta$ with dimension $d$.
This process is orchestrated by the \cgls{gs} and performed iteratively without sharing datasets between satellites.
We assume that the satellites have very limited computational resources available for the solution of \cref{opt}.
Hence, we consider the case where the satellites work on \cref{opt} between visits to the \cgls{gs} and use the contact time to do an exchange of model parameters $\vec\theta$.

\subsection{Federated Learning Background} \label{sec:flbasic}
Solving the \cgls{ml} training problem \cref{opt} distributedly under the assumptions of intermittent connectivity, heterogeneous datasets, and without sharing local raw data is known as \cgls{fl}. The most widely employed approach to this problem is the FedAvg algorithm \cite{McMahan2017}. A \cgls{ps} manages the learning process and keeps a global version of the current model parameters $\vec\theta^i$ to be learned. In each global iteration $i$, the ``epoch'', the \cgls{ps} selects a subset $\mathcal S_i$ of available workers to participate in the next round. It transmits the current version of the global model to the selected workers and then waits for all of them to return their results.

The workers perform one or multiple iterations of minibatch \cgls{sgd} over their local dataset. In particular, in each local epoch the local dataset is partitioned in $\lceil \frac{n_k}{B} \rceil$ random batches $\mathcal B \in\mathscr B$ of size $B = \card{\mathcal B}$ and, for each minibatch, a \cgls{sgd} step is performed with learning rate $\eta$ \cite{McMahan2017,bengio2016}. The loss function is based on \cref{opt} and defined as
\begin{equation} \label{eq:surrogateObj}
	g_{\vec\theta'}(\mathcal B; \vec\theta) = \card{\mathcal B}^{-1} \sum\nolimits_{\vec x\in\mathcal B} f(\vec x; \vec\theta) + \tilde g_{\vec\theta'}(\mathcal B; \vec\theta)
\end{equation}
where $\tilde g$ is an optional regularization term \cite{li2018federated}. Upon termination, the updated local model parameters are transmitted to the \cgls{ps}. The whole procedure is given in \cref{alg:worker}.

\begin{algorithm}[t]
	\caption{Worker \cgls{sgd} Procedure} \label{alg:worker}
	\begin{algorithmic}[1]
		\State Receive $(\vec\theta^i, i)$ from the \cgls{ps}
		\State $\vec\theta_k^{i,0} \gets \vec\theta^i,\quad j \gets 0$
		\While {stopping criterion not met} \label{alg:worker:stop}
			\State $\tilde{\mathcal D}_k \gets $ Randomly shuffle $\mathcal D_k$
			\State $\mathscr B \gets $ Partition $\tilde{\mathcal D}_k$ into minibatches of size $B$
			\For {each batch $\mathcal B\in\mathscr B$}
			\State $\vec\theta_k^{\tau,j+1} \gets \vec\theta_k^{i,j} - \eta \nabla_{\!\vec\theta}\; g_{\vec\theta^{i}}\!(\mathcal B; \vec\theta_k^{i,j})$
			\Comment cf.~\cref{eq:surrogateObj}
			\State $j \gets j+1$
			\EndFor
		\EndWhile
		\State Push $(\vec\theta_k^{i,j}, i)$ to the \cgls{ps}
	\end{algorithmic}
\end{algorithm}

After receiving results from all scheduled workers, the \cgls{ps} aggregates the results into a new version of the global model
\begin{equation} \label{eq:supdate}
	\vec\theta^{i+1} = \sum_{k\in\mathcal S_i} \frac{n_k}{\sum_{k\in\mathcal S_i} n_k} \vec\theta_k^{i},
\end{equation}
where $\vec\theta_k^i$ are the local model parameters of worker $k$, and $\vec\theta^{i+1}$ is the new set of global model parameters. After this aggregation step, the \cgls{ps}
starts the next epoch.

This is known as synchronous \cgls{fl} and can lead to slow convergence speed if the \cgls{ps} has to wait for stragglers. One way to address this problem is to incorporate client updates whenever they arrive in an asynchronous fashion.
Such an algorithm was first published in \cite{Xie2020} under the name FedAsync and is shown to outperform FedAvg in some cases. While the client operation in FedAsync is as in \cref{alg:worker}, the \cgls{ps} operates differently and periodically assigns computing tasks to some workers by transmitting the current version of the global model parameters along with the epoch. Client updates are incorporated asynchronously as they arrive. In particular, the update from client $k$ in epoch $i$ is incorporated as
\begin{equation} \label{eq:faupdate}
	\vec\theta^{i+1} = (1-\alpha) \vec\theta^i + \alpha \vec\theta_k^i
\end{equation}
where the mixing factor $\alpha\in(0, 1)$ determines how much weight is given to incoming client updates. This factor is determined as
$\alpha = \alpha' \cdot s(i - \tau_k)$,
where $\alpha'$ is a fixed base weight, $i$ is the current epoch, $\tau$ is the epoch the worker received the global model, and $s(i)\in(0, 1]$ is a problem-specific staleness function that may be used to reduce the weight given to updates based on older version of the global model. The rationale behind this is that such updates are likely to introduce an error into the solution as the global model parameters have already advanced further towards the solution.

\section{Federated Learning on Satellites} \label{sec:fl}
The \cgls{fl} algorithms discussed in \cref{sec:flbasic} were designed under the premise that device availability is driven by a random process and that parallel communication is possible without significant delay.
However, the satellite scenario is fundamentally different in several aspects: the number of workers is a few magnitudes smaller than in terrestrial applications, devices are always available for computation tasks, but communication is only possible during a small and highly predictable time window. In addition, at each time instant only a very small fraction of workers is within range of communication.

While this scenario is best addressed by an asynchronous \cgls{fl} algorithm, we also consider the synchronous FedAvg algorithm as baseline. We first outline the communication protocol, define the satellite operation, and discuss the application of FedAvg and FedAsync to the satellite scenario. In the next section, we design a novel asynchronous algorithm that leverages the predictable connectivity of satellite communications to implement FedAvg without unnecessary delays.

\subsection{Communication Protocol and Satellite Operation}
Communication is implemented in a client server protocol, where all connections are initiated by the satellite. Whenever the satellite is not working on a communication task, it tries to contact the \cgls{gs}. Hence, communication is either initiated when the \cgls{gs} comes within communication range or directly upon completion of a communication task. Upon connection, satellite $k$ transmits a local model parameter update $\vec\theta_k^i$ if one is available and was not previously sent, where $i$ denotes the current global epoch. Then, the \cgls{gs} updates the global model parameters $(\vec\theta^i, \vec\theta^i_k) \mapsto \vec\theta^{i+1}$ and decides whether satellite $k$ should continue computation. If true, the \cgls{gs} transmits the updated global parameter vector to satellite $k$ and terminates the connection. Otherwise, the connection is terminated and the satellite does not reestablish connection during this pass.

The computation task on the satellite is described in \cref{alg:worker}.
To avoid large deviations from the global model due to asynchronous operation and long delays between \cgls{gs} contacts, $L^2$-regularization on the model parameters is employed \cite{li2018federated}, i.e., the regularization term in \cref{eq:surrogateObj} is chosen as
\begin{equation} \label{eq:fedprox}
	\tilde g_{\vec\theta'}(\mathcal B; \vec\theta) = \frac{\lambda}{2} \norm{\vec\theta - \vec\theta'}^2_2
\end{equation}
with parameter $\lambda$. The stopping criterion in line~\ref{alg:worker:stop} is a fixed number of iterations that should be chosen such that the computation is finished before the \cgls{gs} is visited again.

\subsection{Synchronous Ground Station Operation} \label{sec:fedavg}
We start the discussion of \cgls{gs} operation by adapting FedAvg to the satellite scenario. Recall that
the FedAvg server selects, in epoch $i$, a subset $\mathcal S_i$ of workers to perform updates on the current model $\vec\theta^i$ and then waits for the arrival of \emph{all} scheduled results before updating the model according to \cref{eq:supdate}.
A na\"{i}ve adaption of this algorithm to the satellite scenario is given in \cref{alg:fedavg}.
The main loop runs until convergence is determined in line~\ref{alg:fedavg:stop} by any of the usual criteria, e.g., number of epochs, elapsed wall time, or early stopping \cite[\S 7.8]{bengio2016}.
Workers for the new epoch are selected by the function \textsc{Schedule} and stored in $\mathcal S_i$ and $\mathcal R_i$. The next version of the global model is initialized in line~\ref{alg:fedavg:initsched}.
The role $\mathcal S_i$ and $\mathcal R_i$ becomes apparent in the following lines: $\mathcal S_i$ contains the scheduled workers that have yet to receive the current global model parameters, while $\mathcal R_i$ holds the workers that have not yet returned their model update.
Accordingly, the inner loop in lines~\ref{alg:fedavg:1}--\ref{alg:fedavg:2} runs until both sets are empty.
In line~\ref{alg:fedavg:connect}, the \cgls{gs} waits for any satellite to connect. If it is in $\mathcal S_i$, the current global model is sent, the satellite is removed from $\mathcal S_i$ and the connection is terminated. If not in $\mathcal S_i$ but in $\mathcal R_i$, the \cgls{gs} expects that the satellite transmitted a local model update that is incorporated in the new version of the global model parameters in line~\ref{alg:fedavg:update}. Then, the satellite is removed from $\mathcal R_i$ and the algorithm returns to line~\ref{alg:fedavg:connect}.

\begin{algorithm}[t]
	\caption{Synchronous Ground Station Operation (FedAvg)} \label{alg:fedavg}
	\begin{algorithmic}[1]
		\State \textbf{Initialize} epoch $i = 0$, model $\vec\theta^1$, wall time $t$ \label{alg:fedavg:init}
		\While {stopping criterion not met} \label{alg:fedavg:stop}
			\State $i \gets i + 1$ \label{alg:fedavg:up1}
			\State $\mathcal S_i = \Call{Schedule}{t}$ \label{alg:fedavg:schedule}
			\Comment Predictive scheduling of workers
			\State Initialize $\mathcal R_i = \mathcal S_i$, $\vec\theta^{i+1} = \vec 0$ \label{alg:fedavg:initsched}
			\While {$\mathcal S_i \cup \mathcal R_i \neq \emptyset$} \label{alg:fedavg:1}
				\State Wait for any satellite. Upon connection to satellite $k$: \label{alg:fedavg:connect}
				\If {$k\in\mathcal S_i$} \label{alg:fedavg:transmit}
					\State Transmit $\vec\theta^i$ to satellite $k$ \label{alg:fedavg:transmit1}
					\State $\mathcal S_i \gets \mathcal S_i\setminus\{k\}$ \label{alg:fedavg:transmit2}
				\ElsIf {$k\in\mathcal R_i$} \label{alg:fedavg:receive}
					\State Receive model update $\vec\theta^i_k$ from satellite $k$ \label{alg:fedavg:receive1}
					\State $\vec\theta^{i+1} \gets \vec\theta^{i+1} + \frac{n_k}{n} \vec\theta^i_k$ \label{alg:fedavg:update}
					\State $\mathcal R_i \gets \mathcal R_i\setminus\{k\}$ \label{alg:fedavg:receive2}
				\EndIf \label{alg:fedavg:endif}

				\If {$\mathcal S_i \cup \mathcal R_i \neq \emptyset$}
					\State Terminate connection to satellite $k$
				\EndIf
			\EndWhile \label{alg:fedavg:2}
		\EndWhile
	\end{algorithmic}
\end{algorithm}

The key difference to vanilla FedAvg is that the communication is asynchronous to allow scheduling of satellites not simultaneously visible to the \cgls{gs}, while the update is still computed synchronously. An important observation is that the work loop in lines~\ref{alg:fedavg:1}--\ref{alg:fedavg:2} is blocking, i.e., it waits for all scheduled satellites to connect twice to the \cgls{gs} before starting a new epoch. Assuming the satellite does not finish computation within a single pass, this implies that one epoch takes at least one orbital period on average. However, with multiple satellites scheduled, this time increases.

From an optimization theoretic perspective, \cref{alg:fedavg} is equivalent to FedProx \cite{li2018federated}, a FedAvg variant that uses the regularization in \cref{eq:fedprox}. Convergence follows from \cite[Thm.~4]{li2018federated}.

\subsection{Asynchronous Ground Station Operation}
In contrast to \cref{alg:fedavg}, asynchronous \cgls{fl} operation allows the satellites to work on different versions of the global model, leading to reduced delay.
\Cref{alg:server} outlines the operation of the \cgls{gs}. In line~\ref{alg:server:connect}, it waits for a satellite to connect. Communication is assumed as non-blocking, i.e., communication delay is hidden from \cref{alg:server}. If a model update was received, the epoch is advanced in line~\ref{alg:server:up1} and the global model is updated based on its current version and the received update. If the satellite is scheduled for further computation, the new global model is transmitted in line~\ref{alg:server:schedule3}. The connection is terminated in line~\ref{alg:server:term}.

\begin{algorithm}[t]
	\caption{Asynchronous Ground Station Operation} \label{alg:server}
	\begin{algorithmic}[1]
		\State \textbf{Initialize} epoch $i = 0$, model $\vec\theta^0$, wall time $t$ \label{alg:server:init}
		\Loop \label{alg:server:loop}
			\State Wait for any satellite. Upon connection to satellite $k$: \label{alg:server:connect}

			\If {received model update $(\vec\theta_k^\tau, \tau)$} \label{alg:server:update}
				\State $i \gets i + 1$ \label{alg:server:up1}
				\State $\vec\theta^i \gets $ \Call{ServerUpdate}{$i, \tau, \vec\theta^{i-1}, \vec\theta_k^\tau$} \label{alg:server:up2}
				\If {stopping criterion is met}
					\State Exit loop: Go to line~\ref{alg:server:endloop}
				\EndIf
			\EndIf

			\If {\Call{Schedule}{$k, t$}} \label{alg:server:schedule1}
				\State Transmit $(\vec\theta^i, i)$ to satellite $k$ \label{alg:server:schedule3}
			\EndIf \label{alg:server:schedule2}
			\State Terminate connection to satellite $k$ \label{alg:server:term}
		\EndLoop \label{alg:server:endloop}
	\end{algorithmic}
\end{algorithm}

\subsubsection{FedAsync}
\label{sec:fedasync}
Implementation of the \textsc{ServerUpdate} and \textsc{Schedule} procedures in \cref{alg:server} depends on the \cgls{fl} scheme and the communication scenario. In case of FedAsync (cf.~\cref{sec:flbasic}), \textsc{ServerUpdate} first computes the mixing factor $\alpha$ based on a staleness function $s(i - \tau_k)$ and then returns \cref{eq:faupdate}. As no client update can be fresher than one orbital period, we propose to use a hinged staleness function following the definition in \cite[\S 5.2]{Xie2020}. In particular, let $s(i - \tau_k) = \tilde s(t_i - t_{\tau_k})$ where $t_j$ is the time epoch $j$ was processed at the \cgls{gs} and
\begin{equation} \label{eq:asyncstaleness}
	\tilde s(t) = \begin{cases}
		1 & \mathrm{if}\ t \le (1+\varepsilon) T_{\mathrm{o, max}} \\
		(1 + a (t - (1+\varepsilon) T_{\mathrm{o, max}}))^{-1} & \mathrm{otherwise}
	\end{cases}
\end{equation}
for some small $\varepsilon \ge 0$, a positive constant $a$, and $T_{\mathrm{o, max}}$ being the maximum orbital period within the constellation. The scheduler can easily calculate the value of $s(\cdot)$ at the next pass of a given satellite. Hence, it is possible to conserve energy and computational resources by setting $\textsc{Schedule}(k, t)$ to false if the weight $\alpha$ will be below a certain threshold.

\section{Unrolled Federated Averaging Algorithm}
\label{sec:FedAvgUnrolled}

Synchronous \cgls{fl} procedures applied to ground-assisted satellite in-constellation learning suffer from high latencies. This can be alleviated by asynchronous \cgls{fl} procedures like FedAsync. However, under full client participation and adequately chosen hyperparameters, FedAvg has stronger convergence properties than FedAsync. Hence, it is desirable to implement FedAvg in an asynchronous way. Leveraging on the predictable connectivity of satellites, this is indeed possible.

First, consider a near-polar Walker Delta Pattern Constellation \cite{Walker1984} with single orbital shell
and a \cgls{gs} located at the North Pole. This is a symmetrical scenario where every satellite visits the \cgls{gs} exactly once per orbital period. Moreover, the sequence of connecting satellites to the \cgls{gs} is constant, i.e., if the satellites are ordered such that, within some interval $[t, t+T_o]$ with $T_o$ being the orbital period, the sequence of satellite contacts is $1 \rightarrow 2 \rightarrow 3 \rightarrow \dots \rightarrow K$, then this sequence is repeated in every following orbital period.

In this case, the FedAvg update rule in \cref{eq:supdate} with full client participation can be implemented incrementally without requiring synchronicity in the update phase. In particular, suppose satellite $k$ visits the \cgls{gs} at time $t_{i_1}$ (with epoch $i_1$) and again at $t_{i_2} = t_{i_1} + T_o$. Then, the client update $\vec\theta_k^{i_2}$ is based on $\vec\theta^{i_1+1}$ and can be incorporated in the global model as
\begin{equation} \label{eq:fedavg}
	\vec\theta^{i_2+1} = \vec\theta^{i_2} - \alpha_k (\vec\theta^{i_1}_k - \vec\theta_k^{i_2}),
\end{equation}
where the weight $\alpha_k = \frac{n_k}{n}$ accounts for different dataset sizes.
There are exactly $K$ updates in each orbital period and, due to the periodicity of the satellite-\cgls{gs} contact order, the resulting model after a multiple of $K$ epochs should be close to that from \cref{alg:fedavg}.
More generally, consider a constellation where the revisit period of each satellite with respect to an arbitrarily located (but fixed) \cgls{gs} is approaching the same value. Then, the proposed procedure in \cref{alg:server} with \textsc{ServerUpdate} function as in \cref{eq:fedavg} converges as established in  \cite[\S 2]{Nedic2001}.

Finally, consider a constellation with multiple orbital shells. The assumption about equal revisit rates usually does not hold in this case. Following the discussion in \cite{Nedic2001} this does not prevent convergence but might result in a biased solution. However, this is also the case for many other state-of-the-art \cgls{fl} algorithms, including the methods presented here. Indeed, the numerical results in the next section will show that this effect is much less pronounced in this algorithm than in the asynchronous baseline FedAsync.

\section{Empirical Results} \label{sec:numeval}
We numerically evaluate the performance of the proposed algorithms in terms of the test accuracy on the MNIST \cite{MNIST} and CIFAR-10 \cite{CIFAR} datasets. For MNIST, we train a logistic regression model with 7850 trainable parameters \cite{li2018federated}. The expected accuracy of centralized training is around \SI{89}{\percent}. For CIFAR, we train a ResNet-18 that can achieve an accuracy of slightly above \SI{90}{\percent} when trained centrally \cite{resnet}.
The training dataset is distributed randomly over all workers with equal local dataset sizes. Each satellite operates according to \cref{alg:worker} with $\eta = 0.1$ and $\lambda = 0$, where a single pass over the local dataset is done in batches of size 10 between \cgls{gs} contacts.
We rely on the FedML framework \cite{he2020fedml} for our \cgls{fl} implementation.
In the results, we refer to \cref{alg:fedavg} as ``FedAvg,'' to the asynchronous baseline in \cref{sec:fedasync} as ``FedAsync,'' and to the algorithm in \cref{sec:FedAvgUnrolled} as ``FedSat.'' The staleness function for FedAsync, if used, has parameters $\varepsilon = 0.01 $ and $a = 5 (1+\epsilon)   T_{o,\max}$, where, by Kepler's third law, $T_{o,\max} \approx \SI{127}{\minute}$. The mixing parameter $\alpha$ for FedAsync was fine-tuned for each experiment individually.

A satellite constellation with two orbital shells at altitudes \SI{500}{\km} and \SI{2000}{\km}, respectively, containing five satellites each is considered. Both are Walker Delta constellations \cite{Walker1984} with inclination angle of \SI{80}{\degree} and five orbital planes. They are shifted such that the minimum difference in \gls{raan} between shells is \SI{36}{\degree}. The minimum elevation angle $\alpha_e$ is \SI{10}{\degree}. For the non-IID cases, half of the available classes are distributed to the \SI{500}{\km} orbital shell and the other half to that at \SI{2000}{\km}.

\begin{figure}
	\centering
	\begin{tikzpicture}
		\begin{axis} [
			accuracyplot,
			longtime,
			ymin=0.4,
			yminorgrids = true,
			legend entries = {FedSat, FedAvg, FedAsync, FedAsync $s(t) = 1$}
		]
			\pgfplotstableread[col sep=comma]{CSV_data/Bremen_NIID_MNIST_stale.dat}\tbl

			\addplot table[y=FedSat] {\tbl};
			\addplot table[y=FedAvg] {\tbl};
			\addplot table[y=FedAsync s(t)] {\tbl};
			\addplot+[densely dotted] table[y=FedAsync] {\tbl};
		\end{axis}
	\end{tikzpicture}
	\vspace{-2.5ex}
	\caption{Top-1 accuracy for a \cgls{gs} in Bremen with Non-IID MNIST data.}
	\label{fig:hbNonIID-mnist}
\end{figure}
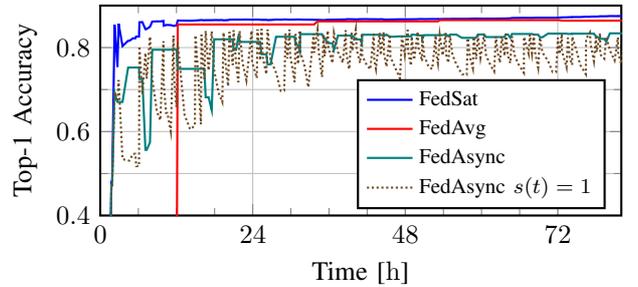

\begin{figure}
	\centering
	\begin{tikzpicture}
		\begin{axis} [
			accuracyplot,
			longtime,
			legend entries = {FedSat, FedAvg, FedAsync, FedAsync $s(t) = 1$},
			legend pos = north west,
			legend columns = 2,
			transpose legend,
			height=.64*\axisdefaultheight,
			legend style = {
				font=\scriptsize,
				inner sep=1pt,
				nodes={inner sep=1pt}
			},
		]

			\pgfplotstableread[col sep=comma]{CSV_data/Bremen_NIID_Update.dat}\tbl

			\addplot table [y=FedSat] {\tbl};
			\addplot table [y=FedAvg] {\tbl};
			\addplot table [y=FedAsync s(t)] {\tbl};
			\addplot table [y=FedAsync] {\tbl};
		\end{axis}
	\end{tikzpicture}
	\vspace{-2.5ex}
	\caption{Top-1 accuracy for a \cgls{gs} in Bremen with Non-IID CIFAR data.}
	\label{fig:hbNonIID-cifar}
	\vspace{-1ex}
\end{figure}
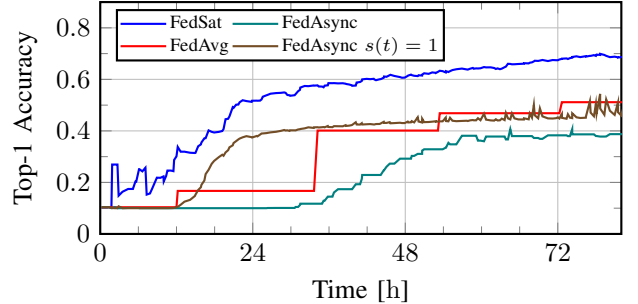

First, consider the case where the \cgls{gs} is located in Bremen, Germany, and the data has non-IID distribution. This scenario poses considerable challenges to the algorithms due to non-uniform device participation and heterogeneous datasets. \Cref{fig:hbNonIID-mnist,fig:hbNonIID-cifar} display the test accuracy for MNIST and CIFAR, respectively.
FedAvg exhibits almost instantaneous convergence for MNIST after a delay of $2 T_{o,\max}$, which is due to the simple model. Instead, in the CIFAR experiment, the accuracy resembles a step function with very slow convergence. This clearly shows the inadequacy of FedAvg, and synchronous algorithms in general, in ground-assisted satellite learning. In both experiments, FedAsync exhibits faster convergence than FedAvg but shows inferior training performance. Interestingly, the staleness function proposed in \cref{eq:asyncstaleness} is necessary for stable convergence for MNIST but has negative impact on CIFAR training.
The mixing factor $\alpha'$ is set to 0.5 and 0.1 for MNIST and CIFAR, respectively, and the learning rate $\eta$ for FedAsync is 0.01.
The proposed FedSat algorithm shows superior training performance, both in convergence speed and final test accuracy.

\tikzsetnextfilename{hbniid-mnist}
\tikzpicturedependsonfile{CSV_data/bremen_noniid.dat}

\tikzsetnextfilename{hbniid-cifar}
\tikzpicturedependsonfile{CSV_data/Bremen_NIID_Update.dat}

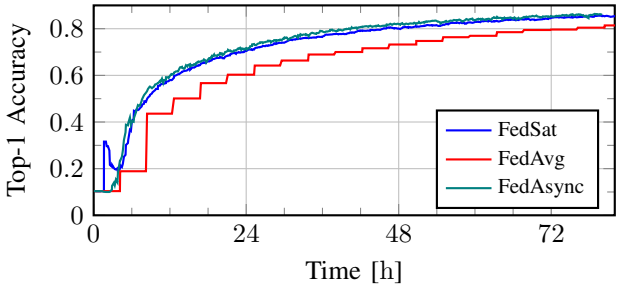
\begin{figure}
	\centering
	\begin{tikzpicture}
		\begin{axis}[
				accuracyplot,
				longtime,
				height=.6*\axisdefaultheight,
				legend entries = {FedSat, FedAvg, FedAsync},
			]
			\pgfplotstableread[col sep=comma]{CSV_data/NP_IID.dat}\tbl

			\addplot table [y=FedSat] {\tbl};
			\addplot table [y=FedAvg] {\tbl};
			\addplot table [y=FedAsync out] {\tbl};
		\end{axis}
	\end{tikzpicture}
	\vspace{-2.5ex}
	\caption{Top-1 accuracy for a \cgls{gs} at the North Pole with IID CIFAR data. FedAsync with $\alpha = 0.3$ and without staleness function.}
	\label{fig:npIID-cifar}
\end{figure}

\begin{figure}
	\centering
	\begin{tikzpicture}
		\begin{axis}[
				accuracyplot,
				longtime,
				height=.63*\axisdefaultheight,
				legend entries = {FedSat, FedAvg, FedAsync},
			]
			\pgfplotstableread[col sep=comma]{CSV_data/Bremen_IID.dat}\tbl

			\addplot table [y=FedSat] {\tbl};
			\addplot table [y=Fedavg] {\tbl};
			\addplot table [y=FedAsync] {\tbl};
		\end{axis}
	\end{tikzpicture}
	\vspace{-2.5ex}
	\caption{Top-1 accuracy for a \cgls{gs} in Bremen with IID CIFAR data. FedAsync with $\alpha = 0.3$ and without staleness function.}
	\label{fig:hbIID-cifar}
	\vspace{-1ex}
\end{figure}
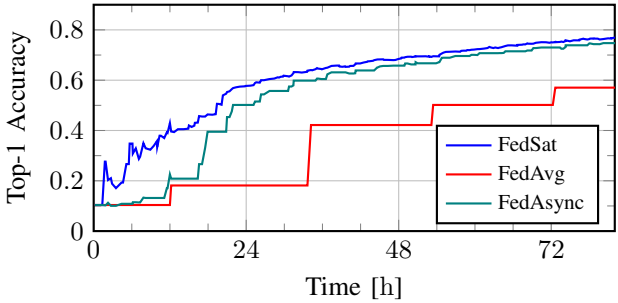

Next, we consider a homogeneous scenario with IID data distribution and \cgls{gs} at the North Pole. The accuracy results are shown in \cref{fig:npIID-cifar}. The FedAvg behavior is as before, but with shorter time periods between updates. This is because the constellation revolves around the \cgls{gs}. Among the asynchronous algorithms, FedAsync has a minor edge over FedSat. However, this requires fine-tuning of the additional hyperparameter $\alpha$, which has optimal value 0.3 in this case.
Finally, \cref{fig:hbIID-cifar} displays results for the same data distribution as before, but for a \cgls{gs} in Bremen. This introduces non-uniform device participation into the previous experiment and could be considered the middle ground between both experiments. The core observation to be made is that FedAsync now performs strictly worse than FedSat. We conclude from this that the proposed method exhibits considerably higher robustness against heterogeneity, which is an important property for the scenario at hand.

\tikzsetnextfilename{npiid-cifar}
\tikzpicturedependsonfile{CSV_data/NP_IID.dat}

\tikzsetnextfilename{hbiid-cifar}
\tikzpicturedependsonfile{CSV_data/NP_IID.dat}

\begin{figure}
	\centering
	\begin{tikzpicture}
		\begin{axis}[
				accuracyplot,
				xmax = 200,
		        xlabel = {Global Epoch},
				legend entries = {FedSat NIID, FedSat IID, FedAsync NIID $\alpha = 0.1$, FedAsync IID $\alpha = 0.3$},
			legend columns = 2,
			transpose legend,
				legend style = {
				font=\scriptsize,
				inner sep=1pt,
				nodes={inner sep=1pt}
			},
			]

			\pgfplotstableread[col sep=comma]{CSV_data/Uniform_settings.dat}\tbl

			\addplot table [y=FedSat NIID] {\tbl};
			\addplot table [y=FedSat IID] {\tbl};
			\addplot table [y=FedAsync NIID] {\tbl};
			\addplot table [y=FedAsync IID] {\tbl};
		\end{axis}
	\end{tikzpicture}
	\vspace{-2.5ex}
	\caption{Top-1 accuracy for uniform client sampling with IID and Non-IID CIFAR data. FedAsync without staleness function.}
	\label{fig:uniform-cifar}
	\vspace{-1ex}
\end{figure}
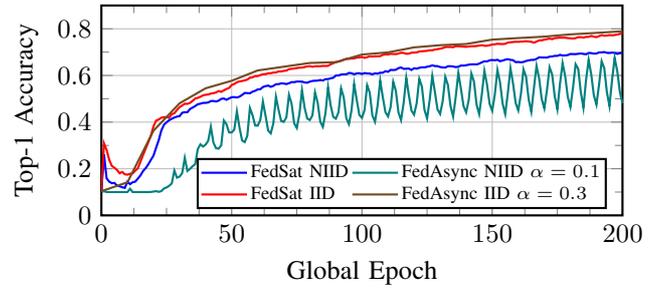

In conclusion, these experiments verify our theoretical considerations. We have observed that a na\"{i}ve implementation of FedAvg \cite{McMahan2017} leads to tremendous delays and that the state-of-the-art in asynchronous \cgls{fl} algorithms, i.e., FedAsync, struggles to deal with the inherent heterogeneity of the satellite learning scenario. We conjecture that this is not only the case for satellite constellations but also for general \cgls{fl} scenarios with heterogeneity. This is supported by an additional simulation in the \cref{fig:uniform-cifar}. Instead, the proposed algorithm shows excellent performance in all experiments.\footnote{The oscillatory behavior of FedAsync in \cref{fig:uniform-cifar} for non-IID data could not be avoided by tuning $\alpha$ and $\eta$. Careful selection of a staleness function and a decaying learning rate might help to dampen this behavior. However, the general trend is apparent and supports our conclusions.}

\tikzsetnextfilename{uniform-cifar}
\tikzpicturedependsonfile{CSV_data/Uniform_settings.dat}

\section{Conclusions}
We have considered \cgls{fl} in \cgls{leo} constellations where satellites collaboratively train a \cgls{ml} model without sharing their local datasets. Unique challenges compared to terrestrial networks were identified and addressed by adapting FedAvg and FedAsync to this setting. We have demonstrated how to unroll FedAvg by exploiting the deterministic worker availability and, effectively, convert it from a synchronous to an asynchronous learning algorithm without sacrificing training performance. This reduces the training time of FedAvg by several hours and leads to an algorithm that outperforms FedAsync both in convergence time and test accuracy. The proposed algorithm also has less hyperparameters to tune than FedAsync.

In this initial work, several topics were left open for future work, including proper scheduling of workers, multiple data exchanges during a single \cgls{gs} pass, and employing multiple \cgls{gs}. These approaches could lead to considerably faster training.


\vspace{-.025ex}

\bibliography{IEEEtrancfg,IEEEabrv,references}

\end{document}

%% file: acronyms.tex
\makeglossaries

\newacronym{raan}{RAAN}{right ascension of the ascending node}
\newacronym{fl}{FL}{federated learning}
\newacronym{ps}{PS}{parameter server}
\newacronym{ml}{ML}{machine learning}
\newacronym{sgd}{SGD}{stochastic gradient descent}
\newacronym{dsgd}{DSGD}{distributed stochastic gradient descent}
\newacronym{isl}{ISL}{inter-satellite link}
\newacronym{gsl}{GSL}{ground to satellite link}
\newacronym{gs}{GS}{ground station}
\newacronym{ecef}{ECEF}{earth-centered, earth-fixed}
\newacronym{eci}{ECI}{Earth-centered inertial}
\newacronym{ofdm}{OFDM}{orthogonal frequency-division multiplexing}
\newacronym{cp}{CP}{cyclic prefix}
\newacronym{los}{LOS}{line-of-sight}
\newacronym{leo}{LEO}{Low Earth Orbit}
\newacronym{eo}{EO}{Earth observation}
\newacronym{iot}{IoT}{Internet of Things}
\newacronym{irs}{IRS}{intelligent reflecting surface}
\newacronym{socp}{SOCP}{second-order cone program}
\newacronym{soc}{SOC}{second-order cone}
\newacronym{dsl}{DSL}{digital subscriber line}
\newacronym{wsee}{WSEE}{weighted sum energy efficiency}
\newacronym{mmwave}{mmWave}{millimeter wave}
\newacronym{dfg}{DFG}{Deutsche Forschungsgemeinschaft}
\newacronym{haec}{HAEC}{Highly Adaptive Energy-Efficient Computing}
\newacronym{hpc}{HPC}{High Performance Computing}
\newacronym{mac}{MAC}{multiple-access channel}
\newacronym{bc}{BC}{broadcast channel}
\newacronym{siso}{SISO}{single-input single-output}
\newacronym{simo}{SIMO}{single-input multiple-output}
\newacronym{miso}{MISO}{multiple-input single-output}
\newacronym{mimo}{MIMO}{multiple-input multiple-output}
\newacronym{af}{AF}{amplify-and-forward}
\newacronym{df}{DF}{decode-and-forward}
\newacronym{cf}{CF}{compress-and-forward}
\newacronym{mwrc}{MWRC}{multi-way relay channel}
\newacronym{dmmwrc}{DM-MWRC}{discrete memoryless multi-way relay channel}
\newacronym{pde}{PDE}{partial data exchange}
\newacronym{fde}{FDE}{full data exchange}
\newacronym{iid}{i.i.d.\@}{independent and identically distributed}
\newacronym{di}{DI} {difference of increasing}
\newacronym{dc}{DC}{difference of convex}
\newacronym{mm}{MM}{mixed monotonic}
\newacronym{mmp}{MMP}{mixed monotonic programming}
\newacronym{awgn}{AWGN}{additive white Gaussian noise}
\newacronym{wgn}{WGN}{white Gaussian noise}
\newacronym{awg}{AWG}{additive white Gaussian}
\newacronym{sic}{SIC}{successive interference cancellation}
\newacronym{snr}{SNR}{signal-to-noise ratio}
\newacronym{sinr}{SINR}{signal to interference plus noise ratio}
\newacronym{inr}{INR}{interference to noise ratio}
\newacronym{zf}{ZF}{zero-forcing}
\newacronym{mrt}{MRT}{maximum ratio transmission}
\newacronym{mmse}{MMSE}{minimum mean square error}
\newacronym{sud}{SUD}{single user decoding}
\newacronym{dof}{DoF}{degrees of freedom}
\newacronym{gdof}{GDoF}{generalized degrees of freedom}
\newacronym{nnc}{NNC}{noisy network coding}
\newacronym{dmn}{DMN}{discrete memoryless network}
\newacronym{csi}{CSI}{channel state information}
\newacronym{pmf}{pmf}{probability mass function}
\newacronym{dmic}{DM-IC}{discrete memoryless interference channel}
\newacronym{ic}{IC}{interference channel}
\newacronym{gic}{GIC}{Gaussian interference channel}
\newacronym{if}{IF}{interference}
\newacronym{ee}{EE}{energy efficiency}
\newacronym{gee}{GEE}{global energy efficiency}
\newacronym{tin}{TIN}{treating interference as noise}
\newacronym{snd}{SND}{simultaneous non-unique decoding}
\newacronym{sd}{SD}{simultaneous decoding}
\newacronym{hk}{HK}{Han-Kobayashi}
\newacronym{rs}{RS}{rate splitting}
\newacronym{rf}{RF}{radio frequency}
\newacronym{pa}{PA}{power amplifier}
\newacronym{lna}{LNA}{low noise amplifier}
\newacronym{lo}{LO}{local oscillator}
\newacronym{adc}{ADC}{analog-to-digital converter}
\newacronym{dac}{DAC}{digital-to-analog converter}
\newacronym{dsp}{DSP}{digital signal processing}
\newacronym{brd}{BRD}{best response dynamics}
\newacronym{br}{BR}{best response}
\newacronym{ne}{NE}{Nash equilibrium}
\newacronym{lhs}{LHS}{left-hand side}
\newacronym{rhs}{RHS}{right-hand side}
\newacronym{ran}{RAN}{radio access network}
\newacronym{qos}{QoS}{Quality of Service}
\newacronym{ngmn}{NGMN}{Next Generation Mobile Networks}
\newacronym{cap}{CAP}{Capacity Adaptation}
\newacronym{bwa}{BW}{Bandwidth Adaptation}
\newacronym{prb}{PRB}{physical resource block}
\newacronym{se}{SE}{spectral efficiency}
\newacronym{tp}{TP}{throughput}
\newacronym{bs}{BS}{base station}
\newacronym{ue}{UE}{user equipment}
\newacronym{mop}{MOP}{multi-objective optimization problem}
\newacronym{gda}{GDA}{generalized Dinkelbach's algorithm}
\newacronym{midcp}{MIDCP}{mixed integer disciplined convex programming}
\newacronym{lp}{LP}{linear program}
\newacronym{brb}{BRB}{branch reduce and bound}
\newacronym{bb}{BB}{branch and bound}
\newacronym{sit}{SIT}{successive incumbent transcending}
\newacronym{oma}{OMA}{orthogonal multiple access}
\newacronym{noma}{NOMA}{non-orthogonal multiple access}
\newacronym{wlog}{w.l.o.g.\@}{without loss of generality}
\newacronym{lsc}{l.s.c.\@}{lower semi-continuous}
\newacronym{usc}{u.s.c.\@}{upper semi-continuous}
\newacronym{kkt}{KKT}{Karush-Kuhn-Tucker}
\newacronym{ptp}{PTP}{point-to-point}